\documentclass[a4paper,twocolumn]{esapub2005} 
\usepackage{times,graphicx}
\usepackage[numbers]{natbib}
\def\lesssim{\mathrel{\hbox{\rlap{\hbox{\lower4pt\hbox{$\sim$}}}\hbox{$<$}}}}

\title{Dynamical processes in the solar radiative interior}
\author{A. Palacios}
\affil{CEA/DSM/DAPNIA/Service d'Astrophysique, CE Saclay b\^at 709,
F-91191 Gif-sur-Yvette (France)}\author{S. Talon}
\affil{D\'epartement de physique, Universit\'e de Montr\'eal, C.P. 6128,
succ. centre-ville, Montr\'eal (Qu\'ebec) H3C 3J7 Canada}\author{S. Turck-Chi\`eze}
\affil{CEA/DSM/DAPNIA/Service d'Astrophysique - CE Saclay b\^at 709 -
F-91191 Gif-sur-Yvette (France)}
\author{C. Charbonnel}
\affil{Observatoire de l'Universit\'e de Gen\`eve, 51 chemin des
Maillettes, CH-1290 Sauverny (Switzerland)}
\affil{LATT- Observatoire Midi-Pyr\'en\'ees, 14 av. E. Belin, F-31400
Toulouse (France)}

\begin{document}


\maketitle

\begin{abstract}
Recent seismic observations coming from acoustic
and gravity modes clearly show that the solar standard model has reached its limits 
and can no longer be used to interpret satisfactorily seismic
observations. In this paper, we present a review of the non-standard
processes that may be added to the solar models in order to improve our
understanding of the helioseismic data. We also present some results
obtained when applying ``non-standard'' stellar evolution to the modelling of
the Sun.  
\end{abstract}

\section{Introduction}

Dynamical processes are thought to occur in stars all across the
Hertzsprung-Russell diagram. The development of advanced and dedicated
observational facilities have greatly improved the constraints that we can
derive from observations. In particular, helioseismology, and more recently
asteroseismology, have given a new insight to our understanding of solar
and stellar physics in both the convective and radiative regions
\citep{Couvidat03}; \citep{TuC04}; \citep{JCD05}. Spectroscopic and
(spectro-)polarimetric data collected for stars of different masses,
metallicities and evolutionary stages, including the Sun, also demonstrate
the actual presence and action of dynamical processes such as rotation and
magnetic fields from direct measurements as well as from indirect evidence
provided by the abundance anomalies.\\ However, while processes such as
rotation, magnetic fields, turbulence and internal waves appear to play a
crucial role in many aspects of stellar evolution, they are not well
understood and their effects are seldom accounted for in stellar and solar
evolution models. At the dawn of a new era for stellar physics, these
processes can no longer be neglected.\\ The dynamical processes
occurring in stellar interiors may have a direct impact on the fundamental
parameters of stars (modification of the basic equations of stellar
structure by rotation or magnetic fields, see for instance
\citep{KT70}). They also lead to transport of matter and/or of angular
momentum. Last but not least, they operate on very different scales at which their action has
to be understood if one wants to properly take them into account in the
stellar modelling. Two complementary approaches have emerged:\\ (1) the
multi-dimensional simulation of the physical process in restricted areas on
dynamical time-scales \citep{BT02};\citep{BBT05}\\ (2) the incorporation of
non-standard\footnote{By non-standard, we mean all processes different from
convection.} physical processes in stellar
evolution codes in order to follow their integrated effects over
evolutionary timescales.\\ In the following, we will adopt the second point
of view and evaluate importance of various physical processes in the
radiative stellar zones. The first approach has been essentially applied to
the convective zones that evolve on dynamical scales, but
few studies of radiative zones also exist (see e.g. \citep{T05})..

\section{Dynamical processes in stellar radiative zones}

To understand the observed stars we need to be able to correctly model
their evolution up to their present status. This implicitly assumes that
the integrated effect of the dynamical processes can be accounted for over
large scales, both spatial and temporal. At present, numerical and
computational restrictions prevent the simultaneous follow up of
evolutionary and dynamical scales, so that we are bound to use
prescriptions for the dynamical processes that can be integrated in a 1D
stellar evolution code.

 {\bf Atomic diffusion}, which consists of gravitational separation,
thermal diffusion and radiative accelerations \citep{AC60},\citep{PM91},
has been the first process added to improve classical stellar evolution
model. The prescriptions for atomic diffusion can be derived from first
principles. In the past decade, the implementation of atomic diffusion
established the new {\em standard} stellar evolution model, which has been
applied with success to various types of stars (Chemically peculiar stars,
horizontal branch stars,...). In particular,
it achieved important improvement of the predicted density and sound speed
profiles in the case of the solar model \citep{BPW95,Brun99,Richard96}.

{\bf Rotation} is a dynamical process at play in all stars. Although the effects
of rotation on stellar structure and evolution have been studied since the
early days of stellar evolution \citep{Eddington25}, it has long been considered a second order
effect and as such it was neglected in stellar computations. In the last decades,
the increasing number of discrepancies between observations and model
predictions have motivated the introduction of rotation and
rotation-induced transport in stellar evolution, as well as the further
development of suitable prescriptions for 1D modelling (see contribution by
S. Mathis, this volume). The main effects of rotation are the modification
of the effective gravity due to the centrifugal forces
\citep{KT70,MM97}, and the transport of angular momentum and
chemical species. In the stellar radiative zones, this transport is ensured
by meridional circulation, which is a consequence of the thermal imbalance
existing in rotating stars, and by turbulence (see \citep{MM00} for more
details).\\ Various approximations and formalisms are currently used in
order to take these effects into account, with two approaches for the
rotational transport of angular momentum. The {\em turbulent diffusion
approach} was first used by \citet{ES78}. It describes the evolution of
angular momentum according to a diffusion equation considering several
linear criteria for hydrodynamical instabilities in order to derive a
turbulent diffusion coefficient (see also \citet{HLW00}). The second
formalism is the {\em meridional circulation approach}, proposed
by \citet{Zahn92}. It describes the angular momentum evolution according
to an advection/diffusion equation. The advection term is related to the
large-scale meridional circulation resulting from the thermal imbalance
existing in a rotating star. The diffusion term is associated with
turbulence generated by only one hydrodynamical source: the shear instability in
its non-linear regime.\\ We use the latter description in its first order
formulation for the transport of angular momentum \citep{Zahn92,MZ98}, and the \citet{CZ92}
formalism to account for the transport of nuclides. In the following, we
will refer to this as the ``{\em rotational mixing of type I}'' (see also
S. Mathis, this volume).\\
Let us note that the coupling of rotation and associated transport
processes with stellar structure results in additional sets of dependent
differential equations that need to be solved at each time step. In the
case of ``{\em rotational mixing of type I}'', the transport of angular
momentum is described by a non-linear system of 5 differential equations,
the resolution of which can be particularly intricate in the dynamical phases of the stellar evolution
(i.e. beyond the main sequence).

 {\bf Magnetic fields} are also observed in a large variety of
stars. Transport processes attached to such fields are particularly
complex, as can be inferred from the intricate field patterns observed at
the surface of solar-type stars. Their incorporation in stellar evolution
models is particularly difficult due to the intrinsic three-dimensional
nature of magnetic fields and the everlasting problem of their
generation. At present, the account for magnetic fields in stellar
structure and evolution consists in treating magnetic perturbations of the
pressure and energy density profiles \citep{LS95}, and/or considering the
magnetic transport of angular momentum and chemical species via
magnetohydrodynamical instabilities
\citep{Spruit02,MM03,MM04,YangBi06}. More recently,
\citet{MZ05a} have also derived a formalism to treat self-consistently the
interplay between meridional circulation, shear-induced turbulence and an
axisymmetric magnetic field (see also S. Mathis, this volume).\\ The
introduction of magnetic fields in stellar evolution codes implies an
additional increase of the number of equations to be solved at each time
step, as well as the addition of terms (Lorentz force) in the equation for
the evolution of angular momentum.

{\bf Internal gravity waves} are the travelling counterpart of the standing
gravity modes of helioseismology (g-modes). They are excited at the base of
the convective zone by Reynolds stresses and/or convective plumes, but the
efficiency of each mechanism is not yet clearly assessed. They transport
angular momentum with a null net contribution if the flux of prograde and
retrograde waves is the same. In a differentially rotating medium however,
the filtering of one of the families of waves can result in a non-null net transport
of angular momentum in the region where they fade away. The spectrum of
internal gravity waves strongly depends on the structure of the stellar convective envelope, and
\citet{TC03} showed that the angular momentum transport by internal gravity
waves is important in the Sun and in main sequence stars with $T_{\rm eff} \lesssim
6700~{\rm K}$ (see S. Talon, this volume for details).\\ At present, the full spectrum
of internal gravity waves cannot be computed in a stellar
evolution code. It needs to be determined in an independent module using
predictions of the stellar evolution models as inputs (extension of the
convective zone, pressure height scales, ...). The feedback on stellar
evolution appears via additional terms in the equations for the transport
of angular momentum and chemicals solved in the rotating case \citep{TC05}.

\section{Dynamical models of the Sun}

In the previous section, we briefly reviewed the main dynamical
processes that will influence the stellar structure and evolution. Rotating
models have proved to largely improve our understanding of massive stars
\citep{HLW00} (see also \citet{MM00} and references therein).\\ In
low-mass stars the picture is somewhat more complex. In these stars, the
surface convective zone is much deeper than for more massive objects, and
dynamical processes such as magnetic transport and internal gravity waves
may become important. Their interaction with meridional circulation and
turbulence must be accounted for.\\ Observational evidence for non-standard transport
processes of both angular momentum and chemicals exist in the case of
late-type stars, and the most striking ones are the lithium abundances and
the helioseismic data. Main sequence F-type stars with $T_{\rm eff}\in
\left[6300~{\rm K}; 7100~{\rm K}\right]$ exhibit lithium depletion at their
surface, and lie in the so-called ``lithium dip''
\citep{BoesgaardTripicco86,Balachandran95}. On the other hand, sharp
transition in rotational velocities and stellar activity also occur in the
range of effective temperature of the Li dip, suggesting a strong interplay
between Li abundances and the action of dynamical processes.\\
The other key observation is the helioseismic rotation inversion which reveals
that, below the convective zone, differential rotation in latitude
disappears and that the solar radiative zone rotates more or less as a solid body down
to at least $r \simeq 0.3~R_\odot$ (see Garcia et al., this
volume). This suggests that efficient transport processes of angular
momentum are at play in the solar radiative interior.\\ In short, on the
main sequence, Pop I low-mass stars experience efficient enough transport of angular momentum so
as to flatten out the rotation profile in the solar interior, but mild
transport of chemicals that allows for moderate destruction of lithium in
their convective envelope.\\ In the following, we present results for dynamical solar models,
and illustrate how stellar evolution together with observational
constraints can provide clues for the understanding of dynamical processes
along the evolutionary path of stars.

\subsection{Rotation}
\begin{figure}
\centering
\includegraphics[width=0.8\linewidth,angle=270]{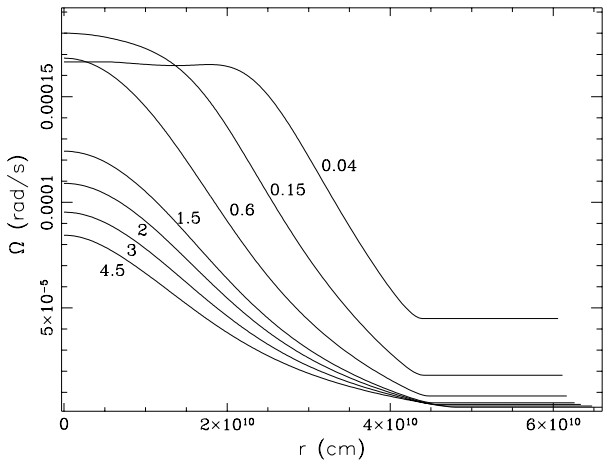}
\caption{Evolution of the angular velocity profile inside the Sun for
  transport of angular momentum by meridional circulation and shear-induced
  mixing according to Zahn (1992). The labels on the curves indicate the
  age in Gy. {\it Reproduced from Talon \citep{Talon97}}.\label{fig:0}}
\end{figure}

\begin{figure}
\centering
\includegraphics[width=0.8\linewidth,angle=-90]{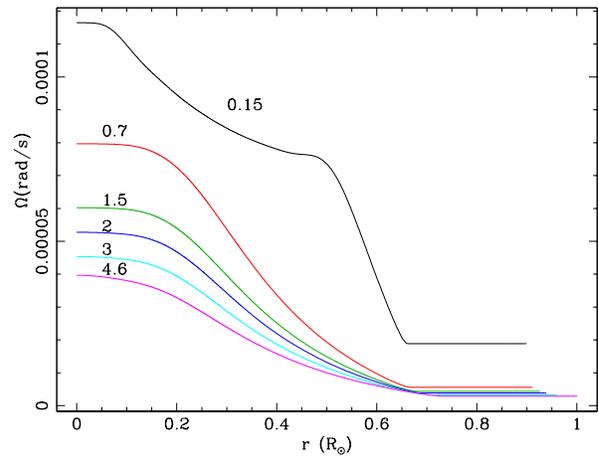}
\caption{Angular velocity profile in the rotating solar model at 0.15, 0.7,
  1.5, 2, 3 and 4.6~Gy.\label{fig:4}}
\end{figure}

\begin{table*}
  \begin{center}
    \caption{Characteristics of our calibrated solar models at 4.6
    Gyrs. }\vspace{1em}
    \begin{tabular}[b]{lrcc}
      \hline
      Parameter & Classical model & Standard  model & Rotating model \\
      \hline
      Atomic diffusion  & no & yes & yes \\
      $\upsilon_{\rm ini}$  & 0 & 0 & $50~{\rm km.s^{-1}}$ \\
      $\upsilon_\odot$  & 0 & 0 & $2~{\rm km.s^{-1}}$ \\
      $\Delta$R/R$_\odot$  & 4.17 10$^{-6}$& 1.2 10$^{-6}$ & 2 10$^{-6}$ \\
      $\Delta$L/L$_\odot$ & 2.35 10$^{-6}$  & 4.8 10$^{-5}$ & 4.06 10$^{-4}$ \\
      $\alpha$ & 1.6746 & 1.9280 & 1.6042 \\
      Y$_0$ & 0.27188 & 0.28501 & 0.26668 \\
      Z$_0$ & 0.01740  & 0.02068 & 0.01620 \\
      (Z/X)$_0$ & 0.02448 & 0.02978 & 0.02259 \\
      Y$_s$ & 0.2719 & 0.2521 &0.2686 \\
      Z$_s$ & 0.01749 &  0.01789 & 0.0175\\
      (Z/X)$_s$ &  0.02461 & 0.02451 & 0.0245\\
      Y$_c$ & 0.62164 & 0.63202 & 0.60173\\
      Z$_c$ & 0.01793 & 0.02389 & 0.01669 \\
      T$_c$ x $10^6$ (K) & 15.48 & 15.80 & 15.33\\
      $\rho_c$ (g.cm$^{-3}$) & 1.511 10$^2$ & 1.526 10$^2$ & 1.467 10$^2$\\
      R$_{\rm BCZ}$ (R$_\odot$) & 0.7302 & 0.7067 & 0.7405\\
      \hline \\
      \end{tabular}
    \label{tab:table1}
  \end{center}
\end{table*}

The first rotating models that were applied to the solar case used the {\em
turbulent diffusion approach}. In their early work, \citet{Pinsonneault89}
explored the parameter space of their formalism and adjusted the 6 main
parameters to observational constraints from both the present Sun and other
stars that can be used to trace the young Sun (pre-main sequence and early
main sequence) and the Sun to come (sub-giant). Among these parameters,
three were arbitrary factors that allow the fine tuning of\\ (1) the
transport of angular momentum with respect to chemicals efficiency ($f_c$),\\
(2) the inhibiting effect of mean molecular weight gradients on the angular
momentum evolution ($f_\mu$),\\ (3) and the efficiency of angular momentum
transport by hydrodynamical turbulence ($f_\omega$).\\ The authors could find
a combination of parameters that reproduced the present Sun in terms of
luminosity, radius, surface rotation rate and lithium abundance. However,
they obtained a considerable amount of differential rotation in the inner 60\%
in radius of their model. Last but not least underline that they had to reduce by a
factor 20 the efficiency of nuclides transport in order for their
model to have the required lithium depletion. Updated versions of this
work was published by \citet{CDP95}, yielding very similar results.\\ Later
on during the 90's, models using a simplified version of the {\em
meridional circulation approach} were computed by
\citet{Richard96}. In that work, the transport of chemicals was solved for
atomic diffusion and rotation-induced mixing. The transport of angular
momentum was not fully solved, but considering the asymptotic regime of
the moderate wind case and given solid-body rotation, the authors could
derive an estimate of the meridional circulation velocity and of the
associated effective diffusion coefficient for the equation for the
transport of chemicals.\\ The best model that was obtained with this
approach included transport of chemicals by atomic diffusion and simplified
rotation-induced mixing, and fitted nicely the present Sun in terms of
radius, luminosity, lithium and beryllium surface abundances.\\ It was however
lacking consistency, not only because angular momentum evolution was not followed
, but also because of tuned mean molecular weight feedback on the
chemicals transport.\\ \citet{MZ97} and \citet{Talon97} applied for the
first time the {\em rotational mixing of type I} as described in
\citep{Zahn92} in order to evaluate the transport of angular momentum in
the Sun. Assuming a solid-body rotation on the ZAMS and associated initial
equatorial velocity of $100 {\rm km.s^{-1}}$, and magnetic torquing at the
surface in order to have $\upsilon_{\rm surf,\odot} = 2 {\rm km.s{-1}}$, they
found that a steep gradient of angular velocity develops in the solar
interior and remains at the age of the Sun. The rotation profiles that they
obtained are reproduced in Fig.~\ref{fig:0} from \citet{Talon97}. 

In the following, we adopt a similar approach, and present new results for
the rotating Sun, obtained with the best available formulation for the
transport of angular momentum within the framework of 1D stellar
evolution. We have implemented the {\em rotational mixing of type I} in the
STAREVOL V2.81 stellar evolution code. The reader is refereed to
\citet{PTCF03,PCTS06} for a description of standard and non-standard
inputs. We have implemented an update of the equation of state and
currently use the FreeEOS-2.0.0 package by Allan Irwin\footnote{The
FreeEOS-2.0.0 package is the latest release of a FORTRAN source code
developed by A. Irwin, and that is made publicly available at {\em
http://freeeos.sourceforge.net} under the GNU General Public License.}. We
used the solar chemical composition by \citet{GN93}, but the differential
effects presented here would be little affected by the use of the new solar
chemical composition. The reader is referred to N. Grevesse's and
J. Guzik's contributions in this volume for more details concerning the new
solar abundances.\\ 
\begin{figure*}[ht]
\centering
\includegraphics[width=0.4\linewidth]{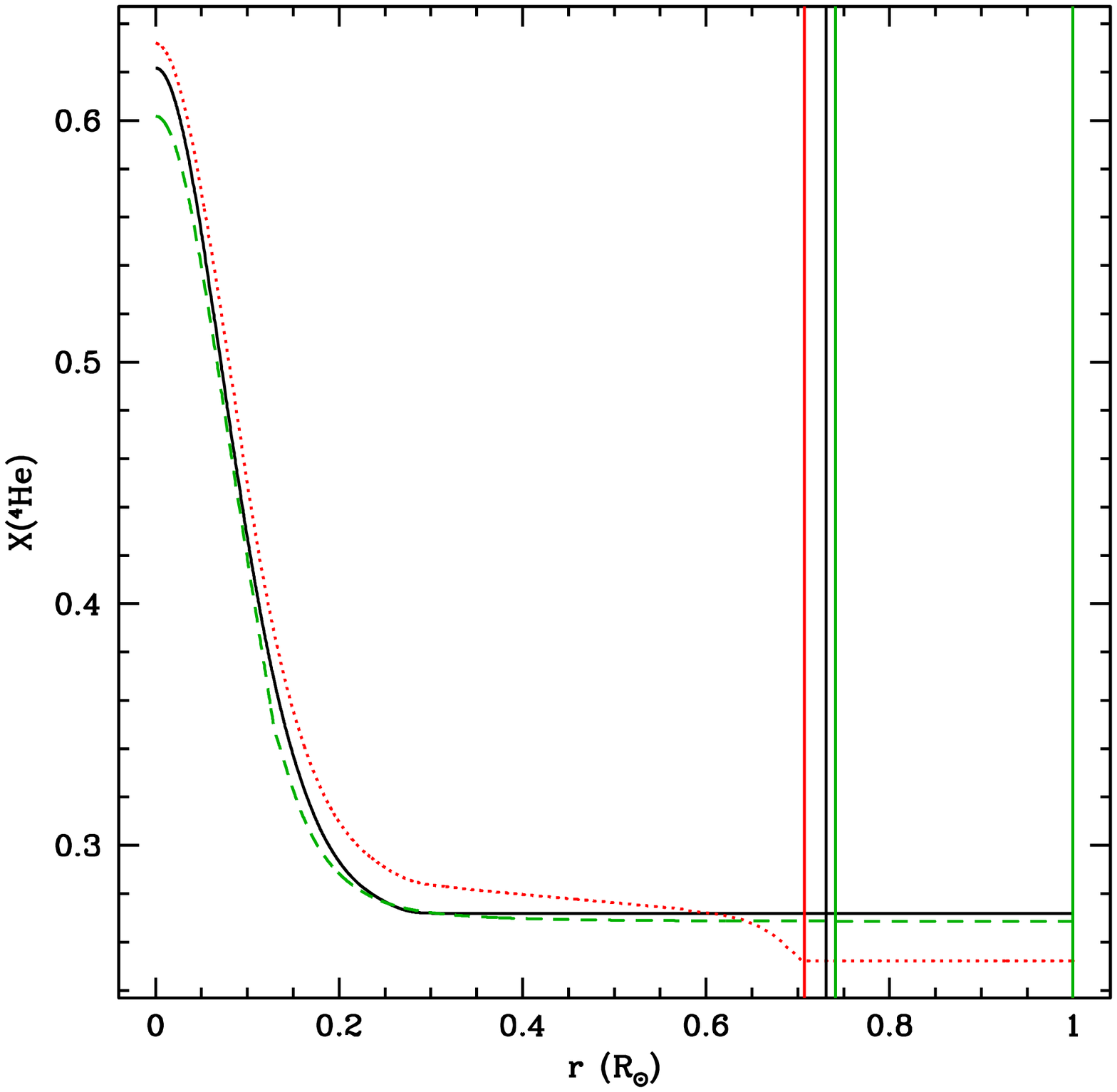}%
\includegraphics[width=0.4\linewidth]{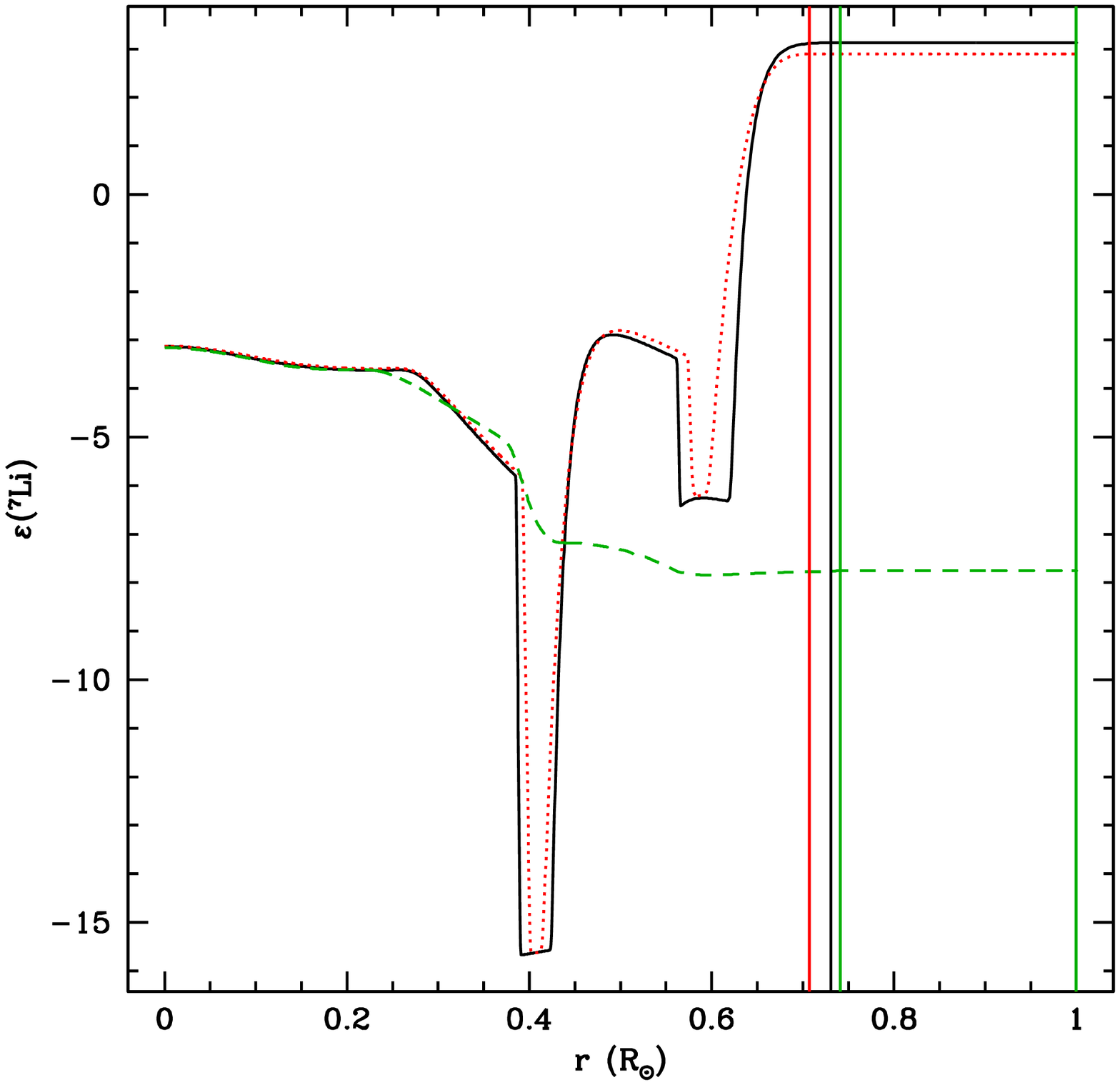}
\caption{Helium mass fraction and lithium abundance profiles in three
  calibrated solar models. Black solid lines, red dotted lines and green dashed lines represent results
  for the standard model, the classical model and the rotating model
  respectively. Vertical solid lines indicate the limits of the surface
  convective zone in each model (from left to right: classical model,
  standard model and rotating model).\label{fig:2}}
\end{figure*}
The characteristics of the models presented here are given in Table 1. We
adopted the following solar values : $R_\odot = 6.9599 \times 10^{10}~{\rm
cm}$, $L_\odot = 3.846 \times 10^{33}~{\rm erg.s^{-1}}$, $Z/X_\odot =
0.0245$. Considering the Sun as a typical G star stars, it might have
undergone strong magnetic braking in its early years on the main
sequence. We follow Kawaler \citep{Kawaler88} and use a braking law of the
form $dJ/dt \propto \Omega^3$. We start with a solid-body rotation on ZAMS,
with an initial rotational velocity typical of late-type stars of $50~{\rm
km.s^{-1}}$. We then let the angular momentum evolve under the combined
action of meridional circulation and turbulence in the radiative region
(the convective region are assumed to have rigid rotation), so that its
surface velocity reaches $2~{\rm km.s^{-1}}$ at 4.6~Gy.\\ Table~1 presents
the main characteristics of three models: a {\em classical} model, with no
rotation and no atomic diffusion, a {\em standard} model, with atomic
diffusion only, and a rotating model, that also includes atomic diffusion. All
models reproduce the radius and luminosity of the present Sun, as well as
the metal fraction for the \citet{GN93} chemical composition. In addition,
the {\em standard} model leads to a radius at the base of the convective
envelope that is compatible with helioseismology.\\ The rotating model
presents some striking features that are worth commenting with further
details.\\ Figure~\ref{fig:4} shows the evolution of the angular velocity
profile inside our calibrated rotating Sun at different ages on the main
sequence. This figure is very similar to Fig.~\ref{fig:0} : during the
first 600~My spent on the main sequence, the large torque applied at the
surface is very efficient at slowing down the surface convective zone. In
the radiative zone, this creates a large gradient of angular velocity. As evolution
proceeds, the profile changes slowly, and we can see that a fraction of
angular momentum is extracted from the core.  However, advection by
meridional flow dominates over the turbulent diffusion, and since it is not
an efficient process, so that at the age of the Sun, the gradient of
angular velocity is still very strong.\\ This result confirms the previous
calculations. The strong gradient of angular momentum is mainly built up
due to the angular momentum losses at the surface. 
The strong extraction of angular momentum in the Sun is related to the fact that in solar-type
stars, the convective envelope is deep enough so as to allow the magnetic
torque to get rooted in it and to be more efficient. This is constrained by
the evolution of rotation velocities of solar-type stars in open clusters
(see \citet{TC04}).\\ Let's now examine the consequences of such a large
$\Omega$-gradient on the internal stratification of the model. Table~1
indicates that, contrary to the {\em standard} solar model, in the rotating
model the surface helium mass fraction increases as the model evolves on
the main sequence. The large $\Omega$-gradient in the radiative interior
allows indeed the shear instability to develop and become turbulent. The
associated diffusion coefficient is large ($D \simeq 10^2 - 10^4~{\rm
cm^2.s^{-1}}$) and more efficient than atomic diffusion, which is dominated
by element settling in the Sun. As diffusion is a ``down gradient''
process, $^4{\rm He}$ is transported from the inner to the outer
regions. When the model reaches 4.6~Gy, the helium gradient is smaller in
the radiative zone, and the gradient that would have been built by element
settling below the convective zone, has been smoothed out (see
Fig.~\ref{fig:2}). The lithium abundance profile has a positive slope, and
shear-induced turbulent diffusion, that dominates the transport of
chemicals, induces inward transport down to regions where lithium is easily
burnt. As a result of this efficient mixing, lithium is destroyed in the
envelope and our rotating model is devoid of Li at 4.6~Gy.\\ The diffusion
of chemicals induces a modification of the opacity in the radiative zone
that translates into shallower convective zone, as mentioned above.

\begin{figure}
\centering
\includegraphics[width=0.8\linewidth,angle=-90]{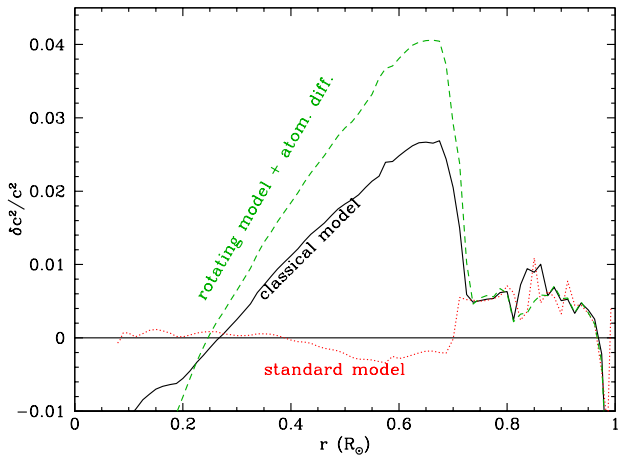}
\caption{Relative difference in sound speed between the Sun and the three
  calibrated solar models as indicated on the plot.\label{fig:1}}
\end{figure}

\begin{figure}
\centering
\includegraphics[width=0.8\linewidth,angle=-90]{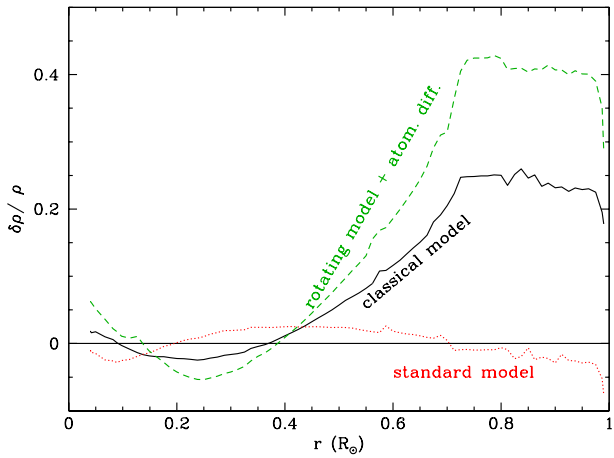}
\caption{Relative difference in density between the Sun and the three
  calibrated solar models as indicated on the plot.\label{fig:3}}
\end{figure}

Figures~\ref{fig:1} and~\ref{fig:3} present respectively the comparison
between the predicted sound speed and density, and the profiles derived
from helioseismic inversions. The differences between the actual Sun and
the {\em standard} model are very small, mainly because of the helium
gradient that appears in this model due to efficient settling below the
convective zone (see Fig.~\ref{fig:2}). On the other hand, the rotating
model predictions largely differ from the helioseismic profiles in the
radiative region, where efficient transport of chemicals occurs. Below the
convective zone, the peak in the $\delta c^2/c^2$ profile is larger than
that in the {\em classical} model, and the mismatch is also larger in the
inner regions. The diffusion of hydrogen and helium ($^3{\rm He}$ and
$^4{\rm He}$) modifies the nuclear energy generation, and the opacity. We
find the same result as \citet{Brun98}, that is to
say that diffusion of the $^3{\rm He}$, which participates in the major
reactions of the {\em pp} chains, increases the discrepancy between
computed and inverted sound speed.\\ Concerning these figures, let us
finally note here that in these first applications of the STAREVOL code to the
solar case, we obtain a mismatch of the sound speed in the convective
region that remains to be explained. Similarly, the obtained differences
for the {\em classical} model appear to be large compared to what is
commonly published for the Sun, but this should not affect the differences
that between the rotating and the non-rotating models that we have discussed.

The rotation and the surface abundances of light elements in the Sun cannot
be reproduced by rotation-induced mixing alone. This provides evidence that
additional transport processes act together with wind-driven meridian
circulation, to extract more angular momentum. Additional extraction of
angular momentum will result in lower differential rotation, and this will
inhibit the shear-induced transport of chemicals.\\ The internal structure
of the Sun provides some clues for the possible additional processes that
could achieve the extra transport of angular momentum. We mentioned in the
above that magnetic torquing is stronger in this region of the
Hertzsprung-Russell diagram due to the extent of the convective zone,
suggesting that magnetic fields could play a non-negligible for the
transport of angular momentum in solar-type stars. On the other hand, the
flux of internal gravity waves becomes important for main sequence stars
with effective temperatures cooler than $T_{\rm eff}<6700~{\rm K}$, and net transport
of angular momentum is expected in the radiative zone.\\ In the last couple
of years, both of these processes have been investigated, leading to
promising results.

\subsection{Rotation and internal gravity waves}

\begin{figure}
\centering
\includegraphics[width=0.8\linewidth]{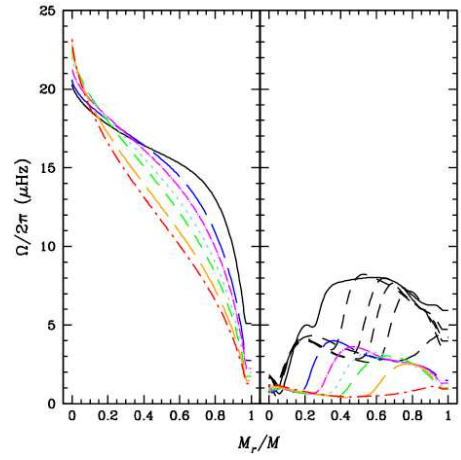}
\caption{Evolution of the angular velocity profile inside a model of 1
  ${\rm M}_\odot$ in presence of rotational mixing of type I (left panel)
  and rotational mixing + internal gravity waves (right panel). The
  long-dashed-dotted lines represent the profiles at 4.6~Gy.{\it Reproduced
  from \cite{CT05}.}\label{fig:6}}
\end{figure}

After the first studies by \citet{Talon97} and \citet{KTZ99} on the
potential of internal gravity waves generated by Reynolds stresses at the
base of the solar convective zone, \citet{TC03} re-investigated their
interplay with rotation in stars within the lithium dip (see \S~2). They
computed the flux of momentum associated with the waves, depending on the
position of the star in the dip, and found a strong correlation between
this flux and the depth of the surface convective region. In the cooler
stars (with the lower mass), where lithium is observed to be less depleted,
they found larger flux of momentum than in hotter stars, laying in the hot
side of the lithium dip, where the flux is almost null. In these hotter
stars, the {\em rotational mixing of type I} has been shown to reproduce
simultaneously surface velocities, lithium and beryllium abundances
\citep{TC98,PTCF03}, and waves were not expected to be significant for
the transport of angular momentum. On the contrary, the stars in the cool
side of the dip, with effective temperatures and masses close to the solar
values, could not be explained via the {\em rotational mixing of type I}
alone, because of the lack of additional processes to extract angular
momentum. In \citep{TC03}, we showed that internal gravity waves had the
good properties to account for this extra-extraction of momentum. In
solar-type stars, the large flux of angular momentum associated with the
internal gravity waves combines with the filtering of prograde waves at the
base of the convective envelope, and leads to a net extraction of angular
momentum from the radiative interior. Let us stress that the actual process
is efficient only because differential rotation of the radiative zone
underlying the convective region affect the shear oscillation layer so as
to filter the prograde waves.\\ The Sun having a similar structure as the
Pop~I stars on the cool side of the lithium dip, we computed a model of
$1~M_\odot$ taking into account the transport of angular momentum by
meridional circulation, shear-induced turbulence and internal gravity waves
\citep{CT05}. An additional contribution to the transport of chemicals in
the oscillating shear layer (see S. Talon, this volume) was also added
below the convective region. As in the rotating model presented in the
previous section, we considered a solid-body rotation with an initial
equatorial velocity of $\upsilon_{\rm ZAMS}=50~{\rm km.s^{-1}}$. Let us
stress that this is not a calibrated solar model, but that this will not
affect the behaviour we find. We will present results for a calibrated
solar model in a forthcoming paper (\citet{PTCTC07}). The evolution of the
internal rotation profile that we obtain is presented in Fig.~\ref{fig:6}.
We can see that extraction of angular momentum by internal gravity waves
occurs in several stages (right panel). During the first stage, (black
solid lines) the bulk of the angular momentum is extracted from the core,
which is decelerated. In the following successive stages, internal gravity
waves complete the flattening of the angular velocity profile, until it is
almost flat when the model reaches 4.6~Gy. The extraction is less efficient
as the star evolves since the differential rotation, which causes the
filtering of prograde waves and allows for negative deposition of angular
momentum by the retrograde waves, is less important.  At the same time, the
effect of meridional circulation and shear-induced turbulence is attenuated
in the presence of internal gravity waves. The lithium abundance of the
model decreases much less than what we presented in Fig.~\ref{fig:2}, and
we get the right amount of depletion at the age of the Sun.

Despite the uncertainties on the flux of internal gravity waves, due to
the uncertainties on the excitation mechanism and mixing-length description
of convection, the transport by internal gravity waves is very 
appealing, and certainly does play a major role in main sequence low-mass
stars. It also  gives a consistent framework to explain the evolution of angular momentum
and chemicals in stars of all masses \citep{TC03,TC04,CT05}.

\subsection{Rotation and magnetic fields}

The effects of magnetic torquing on the transport of chemicals and angular
momentum have been studied in detail by \citet{CMcG93,BCMcG99}, who managed
to reproduce the internal solar rotation profile, and underlined the
importance of the field configuration on the derived surface abundance
patterns. They did not take into account the possible dynamical instability
of the field in their approach, that were recently investigated by
\citet{Spruit02}. In the Tayler-Spruit dynamo theory, azimuthal magnetic
fields can be generated in the radiative interior of differentially
rotating stars. The wound up of the toroidal field by the differential
rotation leads to a magneto-hydrodynamical instability that contributes to
the transport of angular momentum and chemicals. The formalism first
derived by \citet{Spruit02} was revised by \citet{MM04} and recently
applied to the solar case by \citet{Eggenberger05}. They consider an
initial solid-body rotation, with a surface equatorial velocity of $50~{\rm
  km.s^{-1}}$ , as we did in the rotating model presented in the previous
section, and end up with a flat rotation profile at the age of the Sun, in
very good agreement with helioseismology. \citet{YangBi06} also computed
solar model with rotation and magnetic transport of angular momentum and
chemicals. They use the {\em turbulent diffusion approach} for the
transport of angular momentum instead of the {\em meridional
circulation}, and also obtain a flatter angular rotation profile when
accounting for magnetic instabilities. In terms of internal rotation, the
difference with non-magnetic models is however not as striking as in
\citep{Eggenberger05}.\\ These recent computations are promising and show
that magnetic fields are a well-behaved candidate for the additional
transport of angular momentum required in the Sun in order to explain its
internal rotation profile. However they rely on still
controversial prescriptions (the Tayler-Spruit dynamo theory destroys the
agreement between rotating models and observations in the case of massive
stars \citep{MM05}), and should be regarded with the appropriated caution.

\section{Towards a dynamical vision of the Hertzsprung-Russell diagram}

In this paper, we have reviewed the main advances in stellar evolution
modelling concerning the implementation of dynamical processes. Rotation
and associated transport processes has been applied with success to a large
number of stars. It is important to note that in the case of {\em
meridional circulation approach}, where transport of angular momentum is
ensured by meridional circulation and shear-induced turbulence, the same
set of input physics describing the rotational transport have been
successfully applied to low-mass and massive stars at different
evolutionary phases. This makes us confident in the prescriptions used in
what we call the {\em rotational mixing of type I}. This formalism is
however essentially successful when applied to fast rotators, and one
encounters some difficulties when trying to apply it to slow rotators.\\ In
the case of the Sun, the tachocline and the flat internal rotation profile
revealed by helioseismic inversions can not be accounted for with this
processes alone. In solar-type stars, we have also seen that lithium
depletion/preservation requires additional processes to counteract the
effect of rotational mixing when the free parameters that describe
turbulence are calibrated on massive stars. Other indirect observations, relying mainly on
abundance anomalies and reconstruction of the global evolution of angular
momentum during the evolution of low-mass and massive stars also provide
clues on the action of additional transport processes and on the weaknesses
of the commonly adopted prescriptions.\\ An outstanding effort has been
done in the last years to understand the physics of magnetic fields and
internal gravity waves, and their interplay with rotation. This effort led
to the development of adapted formalisms as well as direct numerical
simulations, and provides us with reliable tools for the interpretation of
the high quality observational data to come.\\ We saw that, considering the
Sun as a star and using different observational constraints on pre-main
sequence as well as evolved late-type stars to pin down the solar
evolution, has led to great improvement of our understanding of its angular
momentum and surface chemical abundances evolution.\\ The point we want to
make in this contribution is that dialogue between observations, theory and
direct numerical simulations and modelling is a key ingredient to improve
our understanding of stellar physics. The theoretical efforts presented in
these proceedings (see S. Mathis, S. Talon), as well as their application
in models and numerical simulations represent a pool of tools that are
being validated, and will have to be used in the interpretation of the
missions to come.

\bibliographystyle{aa}
\bibliography{soho18}

\begin{thebibliography}{48}
\expandafter\ifx\csname natexlab\endcsname\relax\def\natexlab#1{#1}\fi

\bibitem[{{Aller} \& {Chapman}(1960)}]{AC60}
{Aller}, L.~H. \& {Chapman}, S. 1960, {\em ApJ}, 132, 461

\bibitem[{{Bahcall} {et~al.}(1995){Bahcall}, {Pinsonneault}, \&
  {Wasserburg}}]{BPW95}
{Bahcall}, J.~N., {Pinsonneault}, M.~H., \& {Wasserburg}, G.~J. 1995, {\em
  Reviews of Modern Physics}, 67, 781

\bibitem[{{Balachandran}(1995)}]{Balachandran95}
{Balachandran}, S. 1995, {\em ApJ}, 446, 203

\bibitem[{{Barnes} {et~al.}(1999){Barnes}, {Charbonneau}, \&
  {MacGregor}}]{BCMcG99}
{Barnes}, G., {Charbonneau}, P., \& {MacGregor}, K.~B. 1999, {\em ApJ}, 511,
  466

\bibitem[{{Boesgaard} \& {Tripicco}(1986)}]{BoesgaardTripicco86}
{Boesgaard}, A.~M. \& {Tripicco}, M.~J. 1986, {\em ApJL}, 302, L49

\bibitem[{{Brun}(1998)}]{Brun98}
{Brun}, A. 1998, Ph.D.~Thesis

\bibitem[{{Brun} {et~al.}(2005){Brun}, {Browning}, \& {Toomre}}]{BBT05}
{Brun}, A.~S., {Browning}, M.~K., \& {Toomre}, J. 2005, {\em ApJ}, 629, 461

\bibitem[{{Brun} \& {Toomre}(2002)}]{BT02}
{Brun}, A.~S. \& {Toomre}, J. 2002, {\em ApJ}, 570, 865

\bibitem[{{Brun} {et~al.}(1999){Brun}, {Turck-Chi{\`e}ze}, \& {Zahn}}]{Brun99}
{Brun}, A.~S., {Turck-Chi{\`e}ze}, S., \& {Zahn}, J.~P. 1999, {\em ApJ}, 525,
  1032

\bibitem[{{Chaboyer} {et~al.}(1995){Chaboyer}, {Demarque}, \&
  {Pinsonneault}}]{CDP95}
{Chaboyer}, B., {Demarque}, P., \& {Pinsonneault}, M.~H. 1995, {\em ApJ}, 441,
  865

\bibitem[{{Chaboyer} \& {Zahn}(1992)}]{CZ92}
{Chaboyer}, B. \& {Zahn}, J.-P. 1992, {\em A\&A}, 253, 173

\bibitem[{{Charbonneau} \& {MacGregor}(1993)}]{CMcG93}
{Charbonneau}, P. \& {MacGregor}, K.~B. 1993, {\em ApJ}, 417, 762

\bibitem[{{Charbonnel} \& {Talon}(2005)}]{CT05}
{Charbonnel}, C. \& {Talon}, S. 2005, {\em Science}, 309, 2189

\bibitem[{{Christensen-Dalsgaard} {et~al.}(2005){Christensen-Dalsgaard}, {Di
  Mauro}, {Schlattl}, \& {Weiss}}]{JCD05}
{Christensen-Dalsgaard}, J., {Di Mauro}, M.~P., {Schlattl}, H., \& {Weiss}, A.
  2005, {\em MNRAS}, 356, 587

\bibitem[{{Couvidat} {et~al.}(2003){Couvidat}, {Garc{\'{\i}}a},
  {Turck-Chi{\`e}ze}, {Corbard}, {Henney}, \& {Jim{\'e}nez-Reyes}}]{Couvidat03}
{Couvidat}, S., {Garc{\'{\i}}a}, R.~A., {Turck-Chi{\`e}ze}, S., {et~al.} 2003,
  {\em ApJL}, 597, L77

\bibitem[{{Eddington}(1925)}]{Eddington25}
{Eddington}, A.~S. 1925, {\em The Observatory}, 48, 73

\bibitem[{{Eggenberger} {et~al.}(2005){Eggenberger}, {Maeder}, \&
  {Meynet}}]{Eggenberger05}
{Eggenberger}, P., {Maeder}, A., \& {Meynet}, G. 2005, {\em A\&A}, 440, L9

\bibitem[{{Endal} \& {Sofia}(1978)}]{ES78}
{Endal}, A.~S. \& {Sofia}, S. 1978, {\em ApJ}, 220, 279

\bibitem[{{Grevesse} \& {Noels}(1993)}]{GN93}
{Grevesse}, N. \& {Noels}, A. 1993, Physica Scripta Volume T, 47, 133

\bibitem[{{Heger} {et~al.}(2000){Heger}, {Langer}, \& {Woosley}}]{HLW00}
{Heger}, A., {Langer}, N., \& {Woosley}, S.~E. 2000, {\em A\&A}, 528, 368

\bibitem[{{Kawaler}(1988)}]{Kawaler88}
{Kawaler}, S.~D. 1988, {\em ApJ}, 333, 236

\bibitem[{{Kippenhahn} \& {Thomas}(1970)}]{KT70}
{Kippenhahn}, R. \& {Thomas}, H.-C. 1970, in {\em IAU Colloq. 4: Stellar
  Rotation}, ed. A.~{Slettebak}, 20

\bibitem[{{Kumar} {et~al.}(1999){Kumar}, {Talon}, \& {Zahn}}]{KTZ99}
{Kumar}, P., {Talon}, S., \& {Zahn}, J.-P. 1999, {\em ApJ}, 520, 859

\bibitem[{{Lydon} \& {Sofia}(1995)}]{LS95}
{Lydon}, T.~J. \& {Sofia}, S. 1995, {\em ApJS}, 101, 357

\bibitem[{{Maeder} \& {Meynet}(2000)}]{MM00}
{Maeder}, A. \& {Meynet}, G. 2000, {\em ARA\&A}, 38, 143

\bibitem[{{Maeder} \& {Meynet}(2003)}]{MM03}
{Maeder}, A. \& {Meynet}, G. 2003, {\em A\&A}, 411, 543

\bibitem[{{Maeder} \& {Meynet}(2004)}]{MM04}
{Maeder}, A. \& {Meynet}, G. 2004, {\em A\&A}, 422, 225

\bibitem[{{Maeder} \& {Meynet}(2005)}]{MM05}
{Maeder}, A. \& {Meynet}, G. 2005, {\em A\&A}, 440, 1041

\bibitem[{{Maeder} \& {Zahn}(1998)}]{MZ98}
{Maeder}, A. \& {Zahn}, J.-P. 1998, {\em A\&A}, 334, 1000

\bibitem[{{Mathis} \& {Zahn}(2005)}]{MZ05a}
{Mathis}, S. \& {Zahn}, J.-P. 2005, {\em A\&A}, 440, 653

\bibitem[{{Matias} \& {Zahn}(1997)}]{MZ97}
{Matias}, J. \& {Zahn}, J.~P. 1997, in {\em IAU Symp. 181: Sounding Solar and
  Stellar Interiors - Posters volume}, ed. J.~{Provost} \& F.-X. {Schmider},
  103--104

\bibitem[{{Meynet} \& {Maeder}(1997)}]{MM97}
{Meynet}, G. \& {Maeder}, A. 1997, {\em A\&A}, 321, 465

\bibitem[{{Palacios} {et~al.}(2006){Palacios}, {Charbonnel}, {Talon}, \&
  {Siess}}]{PCTS06}
{Palacios}, A., {Charbonnel}, C., {Talon}, S., \& {Siess}, L. 2006, {\em A\&A},
  453, 261

\bibitem[{{Palacios} {et~al.}(2003){Palacios}, {Talon}, {Charbonnel}, \&
  {Forestini}}]{PTCF03}
{Palacios}, A., {Talon}, S., {Charbonnel}, C., \& {Forestini}, M. 2003, {\em
  A\&A}, 399, 603

\bibitem[{{Palacios} {et~al.}(2007){Palacios}, {Turck-Chi\`eze}, {Talon}, \&
  {Charbonnel}}]{PTCTC07}
{Palacios}, A., {Turck-Chi\`eze}, S., {Talon}, S., \& {Charbonnel}, C. 2007,
  {\em A\&A}, in prep.

\bibitem[{{Pinsonneault} {et~al.}(1989){Pinsonneault}, {Kawaler}, {Sofia}, \&
  {Demarque}}]{Pinsonneault89}
{Pinsonneault}, M.~H., {Kawaler}, S.~D., {Sofia}, S., \& {Demarque}, P. 1989,
  {\em ApJ}, 338, 424

\bibitem[{{Proffitt} \& {Michaud}(1991)}]{PM91}
{Proffitt}, C.~R. \& {Michaud}, G. 1991, {\em ApJ}, 380, 238

\bibitem[{{Richard} {et~al.}(1996){Richard}, {Vauclair}, {Charbonnel}, \&
  {Dziembowski}}]{Richard96}
{Richard}, O., {Vauclair}, S., {Charbonnel}, C., \& {Dziembowski}, W.~A. 1996,
  {\em A\&A}, 312, 1000

\bibitem[{{Spruit}(2002)}]{Spruit02}
{Spruit}, H.~C. 2002, {\em A\&A}, 381, 923

\bibitem[{{Talon}(1997)}]{Talon97}
{Talon}, S. 1997, Ph.D.~Thesis

\bibitem[{{Talon}(2005)}]{T05}
{Talon}, S. 2005, in {\em EAS Publications Series}, ed. G.~{Alecian},
  O.~{Richard}, \& S.~{Vauclair}, 187--196

\bibitem[{{Talon} \& {Charbonnel}(1998)}]{TC98}
{Talon}, S. \& {Charbonnel}, C. 1998, {\em A\&A}, 335, 959

\bibitem[{{Talon} \& {Charbonnel}(2003)}]{TC03}
{Talon}, S. \& {Charbonnel}, C. 2003, {\em A\&A}, 405, 1025

\bibitem[{{Talon} \& {Charbonnel}(2004)}]{TC04}
{Talon}, S. \& {Charbonnel}, C. 2004, {\em A\&A}, 418, 1051

\bibitem[{{Talon} \& {Charbonnel}(2005)}]{TC05}
{Talon}, S. \& {Charbonnel}, C. 2005, {\em A\&A}, 440, 981

\bibitem[{{Turck-Chi{\`e}ze} {et~al.}(2004){Turck-Chi{\`e}ze}, {Couvidat},
  {Piau}, {Ferguson}, {Lambert}, {Ballot}, {Garc{\'{\i}}a}, \&
  {Nghiem}}]{TuC04}
{Turck-Chi{\`e}ze}, S., {Couvidat}, S., {Piau}, L., {et~al.} 2004, {\em
  Physical Review Letters}, 93, 211102

\bibitem[{{Yang} \& {Bi}(2006)}]{YangBi06}
{Yang}, W.~M. \& {Bi}, S.~L. 2006, {\em A\&A}, 449, 1161

\bibitem[{{Zahn}(1992)}]{Zahn92}
{Zahn}, J.-P. 1992, {\em A\&A}, 265, 115

\end{thebibliography}

\end{document}